\documentclass[]{aa}
\usepackage{txfonts}
\usepackage{graphicx}
\usepackage{natbib}
\bibpunct{(}{)}{;}{a}{}{,}
\def\persec{~$\mathrm{s}^{-1}$}

\def\vec#1{\ensuremath{\mathchoice{\mbox{\boldmath$\displaystyle#1$}}
{\mbox{\boldmath$\textstyle#1$}}
{\mbox{\boldmath$\scriptstyle#1$}}
{\mbox{\boldmath$\scriptscriptstyle#1$}}}}

\begin{document}
\title{A new reduction of the raw Hipparcos data}
\author{Floor van Leeuwen\inst{1} \and Elena Fantino\inst{2}% etc
}                     % Do not remove
\offprints{F.van Leeuwen}          % Insert a name or remove this line
\institute{Institute of Astronomy, Madingley Road, Cambridge CB3 0HA, UK \and 
Dipartimento di Astronomia, Universit\`a di Padova, Vicolo dell'Osservatorio 2,
35122 Padova, Italy}
\date{Received: date / Revised version: date}
% The correct dates will be entered by the editor
%
%
\abstract{
We present an outline of a new reduction of the Hipparcos astrometric data,
the justifications of which are described in the accompanying paper. The 
emphasis is on those aspects of the data analysis that are fundamentally 
different from the ones used for the catalogue published in 1997. The new 
reduction uses a dynamical modelling of the satellite's attitude. It 
incorporates provisions for scan-phase discontinuities and hits, most of 
which have now been identified. Solutions for the final along-scan attitude 
(the reconstruction of the satellite's scan phase), the abscissa corrections 
and the instrument model, originally solved simultaneously in the 
great-circle solution, are now de-coupled. This is made possible by starting 
the solution iterations with the astrometric data from the published 
catalogue. The de-coupling removes instabilities that affected  
great-circle solutions for short data sets in the published data. The 
modelling-noise reduction implies smaller systematic errors, which is 
reflected in a reduction of the abscissa-error correlations by about a factor 
40. Special care is taken to ensure that measurements from both fields of view 
contribute significantly to the along-scan attitude solution. This improves 
the overall connectivity of the data and rigidity of the reconstructed sky,
which is of critical importance to the reliability of the astrometric data. 
The changes in the reduction process and the improved understanding of the 
dynamics of the satellite result in considerable formal-error 
reductions for stars brighter than 8th magnitude.
}
\maketitle
\section{Introduction}
\label{intro}
Although the Hipparcos mission \citep{esa97,perry97L,fvl97,koval98} finished 
more than ten, and the reductions some eight, years ago, there are still 
aspects of the mission that are only now becoming fully understood. These 
aspects mostly concern peculiarities of the rotation rates of the satellite 
(and in particular the scan velocity), and methods used to reconstruct these 
rates and the resulting angular displacements as a function of time. This
process, referred to as the attitude reconstruction, ultimately provides
the reference frame against which the scientific products of the mission are 
measured: time-resolved positional information on 118~000 pre-selected stars.
The accuracy of the attitude reconstruction is therefore a determining
factor in the overall quality of the astrometric data produced by the mission,
and, as was stated in the accompanying paper, this process is to a large 
extent determined by our understanding of the rotational motions of the 
spacecraft. It is the progress in this understanding that has played a major
role in the new reduction.

Despite a serious problem with the orbit due to a failure of the apogee boost
motor \citep[see Volume 2 of ][]{esa97,paper1}, 
and the resulting solution instabilities and radiation damage, Hipparcos 
produced results well exceeding the expectations set for the nominal mission. 
This was confirmed by a wide range of statistical tests on the final data 
\citep{areno95,llind95}. At the time of the publication \citep{esa97}, it was 
therefore generally assumed that a complete re-reduction of these data would 
never be attempted or even be needed: it seemed that the best possible 
results had already been obtained by careful merging of the results from 
the two independent reduction chains, FAST \citep{koval92} and NDAC 
\citep{lindeg92}. It was envisaged that any further improvements would be 
done using the published intermediate astrometric data, the abscissa residuals
that were used to derive the astrometric parameters for the program stars 
\citep{vLDWE}.

Full-sky survey missions like Hipparcos are self-cali\-brating: their 
scientific products are also used for defining the instrument characteristics. 
For Hipparcos, the reconstruction of the along-scan attitude uses the same 
measurements that form the input for the astro\-metric-parameter 
determinations. These two reconstructions are separated in two ways: 
measurements at different epochs allow the recognition of displacements 
due to proper motions, while simultaneous measurements in the two fields 
of view (with different along-scan parallax-factors) allow ultimately 
for the recognition of displacements due to parallaxes. However, the process 
of separation is non-linear and the Hipparcos catalogue can 
therefore only be obtained as the result of a sequence of iterations through 
the mission data. The present study can in that sense be understood as a 
further and probably final set of iterations. Starting these iterations with 
the data as presented in the published catalogue for starting values, allows
simplifications in the reductions, which remove potential instabilities 
from the solutions. Furthermore, modern computers make a much faster reduction 
of the data possible, which allows the iterations to take place that are 
necessary to reach photon-count limits on the accuracies for stars as bright 
as magnitude 3 to 4, requiring an accuracy for the reference frame of better
than 0.1~mas. 
 
Since the publication in 1997 doubts have been raised about some of the 
results derived from the catalogue data, most noticeably the distances of 
a few open clusters. The astrometric data for open clusters are based on 
combined results from the cluster members, which, instead of being solved 
for as individual stars, are solved together for a single cluster parallax 
and proper motion \citep{vLDWE,fvl99a,robic99}. The precision of a cluster 
parallax is therefore generally higher than that of individual stars, and 
can indicate in more detail possible problems in the catalogue. In particular
the difference between the expected and observed parallax of the Pleiades 
cluster has been used as an indication that in some difficult areas of the 
sky the mechanisms of the data reductions may not always have been optimal 
\citep{pinso98,pinso00,pinso03,soder98,naray99,reid99}. A likely cause of these
problems was identified by \citet{makar02} and \citet{fvl05} as due to
a problem in the along-scan attitude reconstruction in the presence of
high-density star fields. By not incorporating a correction for the large
weight discrepancies that could occur under these conditions, the along-scan
attitude would be dominated by data from one field of view only, and the 
astrometry for such a high-density field could become (partly) disconnected 
from the rest of the catalogue. This issue will be referred to as the 
connectivity of the data. Good connectivity is one of the two fundamental 
conditions for an astrometric mission of the type of Hipparcos. The other
is a stability requirement on the basic angle between the two fields of
view.

The current study was not initiated with the idea or intention to re-reduce
the Hipparcos data. Instead, between 1998 and 2003, we undertook a study on 
the dynamics of the satellite
as derived from the reconstructed attitude files obtained over the mission
\citep{fantino00,paper3}. The aim was to try to understand the torques acting
on the satellite well enough to use that information in the reconstruction
of the attitude. The uniquely-high accuracy level of the Hipparcos attitude
reconstruction exposes details in external torques never seen before.
However, experiments with the reconstructed attitude for the NDAC
reductions used in the published data showed that, when interpreted as 
rotational velocities and accelerations, details of the reconstructed 
attitude generally contradicted what could be expected for a freely moving 
rigid body. These violations, which are the equivalent of modelling errors, 
reflect in correlations between the abscissa residuals, for which the 
measurements are obtained relative to the reconstructed attitude 
\citep{vLDWE}. We therefore set out to test
the hypothesis that by using a dynamical model for the satellite attitude,
a more reliable reconstruction could be obtained. However, testing this 
hypothesis required re-reducing a large fraction of the Hipparcos raw data. 
The testing exposed a number of defects in the data that could be repaired,
such as the scan-phase jumps and external hits described in the accompanying
paper. Once these defects were taken care off, the attitude noise and error
correlations started to drop dramatically. At the same time the importance of
connectivity in producing absolute parallaxes became much better understood. 
With all that information in hand, the availability of the raw data and a 
new data analysis package, we were left with no other choice than a complete 
re-reduction of the raw data. As the stars that are most affected (those
brighter than 8$^\mathrm{th}$ magnitude) include for example many open 
cluster members and a number of Cepheids, it was judged that a new reduction 
is more than just a technical or data analysis exercise: it is also highly 
desirable from an astrophysical point of view. 

In the original reductions, between 1989 and 1997, it took a considerable 
manpower effort, and six to eight months, to process just once the entire data
stream with the then available hardware. Handling the Hipparcos data has 
become much simpler over the years. The data for the mission had originally
been delivered on 9-track 6250~bpi magnetic tapes, about 1100 in total. In 
the NDAC consortium these had been converted to an early version of the 
optical disk, using a total of about 160 disks. With these disks some random 
access to the data was possible all through the mission. In 1999 these disks 
were replaced by a CD-ROM archive (180 disks), and early 2003 that archive 
was converted to 24 DVDs. The data that used to occupy the space of almost 
an entire office is now kept in the drawer of a desk. With new software 
(written in C$^{++}$, and including extensive display facilities) and new
hardware we are now able to go once through the entire data stream in three 
weeks, and for further iterations in about three to four days. 

We present an overview of the new reduction in Section~\ref{sec:summnew},
with the emphasis on where the new approach is different from the earlier 
analysis. The attitude reconstruction method as used in the new reduction
is described in Section~\ref{sec:att}. The statistics of the new
analysis, demonstrating the internal consistency, are described in 
Section~\ref{sec:intstat}. The derivation of the astrometric parameters is 
described in Section~\ref{sec:astrompar}. Suggestions for external tests, 
to verify formal errors and general conformity of the new results, are 
presented in Section~\ref{sec:extstat}, and our conclusions in 
Section~\ref{sec:concl}.

This, and the accompanying paper, form the culmination of more than 7 years of 
investigations, testing, checking and processing. During that time various 
people have given  advice and on occasion put us back on the right track 
again. In particular advice from Lennart Lindegren, at some critical stages 
of the development, has been highly valuable. Discussions we had on 
intermediate results with Michael Perryman, Rudolf Le Poole, Dafydd Evans, 
Fr{\'e}d{\'e}ric Arenou, Ulrich Bastian and Fran\-\c{c}ois Mignard are also 
much appreciated.

\section{Summary of the new reduction}
\label{sec:summnew} 

The new reduction differs from the earlier reductions by FAST and 
NDAC in two major aspects, both of which concern the reconstruction of the 
along-scan attitude. The first aspect is the dynamical modelling of the
attitude, in which the underlying torques rather than the pointing variations
are modelled, and which constrains the movements of the satellite to the 
physics of a freely rotating rigid body. It was hoped that these physical 
constraints would reduce systematic errors in the attitude. The
improved understanding of the movements of the satellite gained from applying
the dynamic modelling would also be of benefit to the attitude modelling.

The second aspect is the de-coupling of the three parameter sets that were, 
in the original reductions, solved simultaneously in a process referred to 
as the great-circle reduction \citep{vdmar88, vdmar92}. These constituted the
corrections to the abscissae, the along-scan attitude and the instrument 
parameters. The de-coupling has become possible as the 
\textit{a priori} astrometric parameters (the predicted star positions for 
a given circle) are obtained from the published catalogue and have relatively 
low levels of largely white noise. This is quite different from
the starting positions provided by the Input Catalogue \citep{esa89, esa92} 
used in the original reduction. The de-coupling has many advantages: 
\begin{itemize}
\item it prevents the occurrence of instabilities that characterise short 
data sets in the original reduction (see accompanying paper);
\item it no longer uses projection on a reference great circle, which
could introduce additional noise; the sensitivity to errors in the 
reconstruction of the spin-axis position is therefore much reduced;
\item the data now available for determining astrometric parameters are 
field transits rather than the combined transit data over an orbit of the 
satellite; field transits have better defined statistical properties than 
the combined transits for an orbit, referred to as ``orbit transits'', 
and allow for improved detection and elimination 
of disturbed measurements and small-scale grid distortions;
\item the resolution of data used in the three processes can now be adjusted
to the specific requirements, providing in particular an improvement in the
detection of disturbances like scan-phase jumps and satellite hits;
\item it is now possible to apply adjustments for strong weight 
discrepancies between the data in the two fields of view, which is needed to
improve the rigidity of the reconstructed catalogue. 
\end{itemize}
Like the original reductions, the new reduction is a block-iterative
adjustment. The starting conditions for the new reduction, however, reduced
the systematic errors in the \textit{a priori} astrometry to an insignificant 
level, which made it possible to sub-divide one block of the original 
reductions into three independently-solved blocks, greatly improving the
stability of the solutions.

With these changes in place, and a completely re-written software package, 
the statistics of the intermediate reduction results can be followed in 
detail and used to detect at an early stage any problems in the data. 
Improvements are also made by preserving more information in the 
intermediate data products, such as the offset of the sensitive area 
of the main detector (the instantaneous field of view or IFOV), and the
total photon count that contributed to an observation. The nominal signal
modulation parameter $M1$ is also preserved (see the accompanying paper).
The latter two parameters should provide a reliable handle on formal errors 
from the first reductions to the final astrometric parameter determinations.

Some minor changes are applied to the instrument-parameter solution. The
basic angle correction is also included in the corrections for the along-scan
attitude as the systematic difference for the corrections in the two fields
of view. This allows for the occasional correction for basic-angle drifts.
Chromaticity terms, which originally were entered only in the 
sphere solution, can now be solved directly as part of the instrument
parameters. Finally, in the sphere solution corrections are also very 
small. This process too is therefore treated in a differential manner: after 
solving for the new astrometric parameters the remaining residuals are used to 
determine the corrections for the reference phases of each circle as well as
residual colour dependences. Small-scale distortions as 
a function of the transit ordinate can now be investigated too. Provisions
for 6$^\mathrm{th}$ harmonic residual modulations are no longer needed, as 
these originally resulted from instabilities in the great-circle reductions.

The new reduction includes information from the second harmonic of the 
modulated signals from the main grid, similar to that done by the FAST 
consortium. There is, however, a difference in the weights assigned to the
second-harmonic information. Examining the abscissa residuals obtained by
applying various weight ratios for the first and second harmonics shows a
minimum for a significantly lower weight than appears to have been applied 
in the FAST reductions. The weight ratio between first and second harmonic 
in the new reduction is 9 to 1 (which corresponds to the amplitude ratio 
between the first and second harmonic of 3 to 1); in the FAST reduction 
the weight ratio appears to be set at 3 to 1.

The colour variations of large-amplitude red variables are 
incorporated at all layers of the reduction and calibration, using data 
provided by Dimitri Pourbaix \citep{pourb00,knapp01,knapp03,pourb03}. Colour 
variations affect the calibration of the main-grid modulation, the instrument 
parameters and the chromaticity. 

Other aspects of the data reductions are identical or nearly identical to the 
original NDAC or FAST reductions. For example, the reduction of the photon
counts of the main grid follow the NDAC recipe of phase binning, which
had been extensively tested in 1984 on simulated data and performed well all
through the mission. The star mapper data reduction follows the NDAC 
principles of transit recognition and the possibility to fit multiple transits,
but the FAST algorithm for the star mapper background estimation is used.
The instrument-parameter model is implemented in the same way as used by
NDAC, but the constraining of third-order parameters is more like (but not 
identical to) what was implemented by FAST. The use of the gyro data is 
similar to what was done in the NDAC reductions.

\section{The attitude model}
\label{sec:att}
\subsection{Principles of the model}

The attitude model for the new reduction is based on the hypothesis that the
torques affecting the spacecraft during times of observations, both external 
(solar radiation, gravity gradient, magnetic) and internal (the angular 
momentum of the rate-integrating gyros), can be represented as a continuous 
function in time, such as a cubic spline. It further assumes that the spacecraft 
is a free-floating rigid body. Under these assumptions, the inertial rates of 
the satellite are linked to the torques acting on it through the Euler equations:
\begin{equation}
	\mathrm{\textbf{I}}\frac{{\rm d}\vec{\omega}}{{\rm d} t} = \vec{N} - 
		\vec{\omega}\times\mathrm{\textbf{I}}\vec{\omega},
	\label{eq:euler}
\end{equation}
where $\mathrm{\textbf{I}}$ is the calibrated inertia tensor of the satellite, 
$\vec{N}$ the external and internal torques, and $\vec{\omega}$ the 
inertial rates around the satellite axes 
\citep[see for example][]{goldstein80}. Given a model of the torques, the
rates can be derived through integrating Eq.~\ref{eq:euler}.
The pointing direction as a function of time is obtained by integrating the 
inertial rates through an appropriate set of differential equations (depending
on the reference frame for the pointing angles). These two sets of 
integrations require starting points, and 
a new set of starting points is needed each time a discrete interruption of 
the inertial rates or of the pointing occurs. Discrete rate changes took place 
as a result of thruster firings and external hits. The rotational velocity 
changes predicted from the thruster firing lengths are not accurate enough to 
allow for an integration across these firings. The only discrete pointing 
changes that occurred are the scan-phase discontinuities described in the 
accompanying paper. None of these changes affect the continuity of the 
underlying torque model.
 
The principles of the attitude model are maintained from the first rate 
estimates using gyro data to the final along-scan phase estimates using the 
main detector data. The main difference between the different levels of the
attitude reconstruction is the increase in the density of nodes in the
spline fitting, reflecting the increase in accuracy of the data when going
from gyro data to star mapper data, and subsequently to the main-grid
transit data. 

The spline functions used here are exact splines, in which the boundary 
conditions at the nodes have been substituted explicitly. An exact spline 
$f(x)$ of order $n$ with nodes at $x_i$ ($i=1,...,N$) is defined as a 
sequence of polynomials $f_i(x)$ of the kind
\begin{equation}
\label{eq:spline_n}
f_i(x) = \sum_{j=0}^n a_{ij}x^j, \; {\rm with~~} 
x_{i} \le x \le x_{i+1} 
\end{equation}
Boundary conditions at the nodes require exact continuity up to the 
$(n-1)^\mathrm{th}$ 
derivative. Substituting these conditions leaves a central polynomial,
$f_{\bar{i}}(x)$,  and one additional coefficient per node:
\begin{equation}
\label{eq:polyn}
f_i(x) = f_{\bar{i}}(x) + \left\{ \begin{array}{ll} \displaystyle
\sum_{k=\bar{i}+1}^{i} (a_{kn} - a_{(k-1)n})(x-x_{k})^n &
{\rm ~if~} i > \bar{i} \\ \displaystyle
\sum_{k=i}^{\bar{i}-1} (a_{(k+1)n} - a_{kn})(x-x_{k+1})^n & 
{\rm ~if~} i < \bar{i} \\
\end{array} \right. 
\end{equation} 
In this representation, an order-5 spline contains only 2 more degrees of
freedom than an order-3 spline applied to the same interval and using the 
same nodes. The derivatives of a spline function thus defined are again 
exact splines, but of a lower order. A full description and derivation of 
these equations is presented by \citet{paper4}.

Discontinuities in scan-rate or scan phase are accommodated by adding further
zero- and first-order coefficients for each interruption. 

The fitting of the gyro data (inertial rates) is done using an order-4 spline, 
the fitting of positional corrections uses an order-5 spline. Corrections to 
the underlying torques are obtained from the first or second derivatives of 
these functions, and corrections to the starting
points are obtained by evaluating the fit (or its first derivative) at the
reference time for each starting point. These adjustments are non-linear
and require a few iterations before corrections to the underlying torques 
vanish. All these processes, as well as the 
functions used, are described in full detail by \citet{paper4}, where
they are referred to as the Fully-Dynamic-Approach (FDA) for the attitude 
modelling. The calibrations of the external and internal torques, as well
as the inertia tensor, are described by \citet{fantino00,paper3}.

Fitting this model across a short penumbra phase of an eclipse turned out
to be too demanding. Many degrees of freedom have to be added to account
for the variety of penumbra conditions that may be encountered. The penumbra
conditions were affected by the Earth atmosphere (affecting the amount of
solar radiation received as a function of time), changes in the magnetic 
moment of the satellite due to switching the power supply between solar panels
and batteries \citep{paper3} and thermal adjustments of the structure of the 
satellite as a reaction on the sharp temperature changes. 
Short penumbra transitions have therefore been deleted from the data stream. 

External hits and scan-phase discontinuities (jumps) are treated in the same 
way as thruster firings, but only the largest hits are incorporated at all 
stages of the reductions. The effects of small hits and jumps are only taken 
care off in the final stage of the reductions, the along-scan attitude 
fitting. This avoids creating intervals that are too small for the 
star-mapper based attitude fits, while the small hits and jumps only cause 
significant disturbances at the high accuracy level of the final along-scan 
attitude fitting. 

\subsection{Non-rigidity}
\begin{figure}
\centerline{\includegraphics[width=8.5cm]{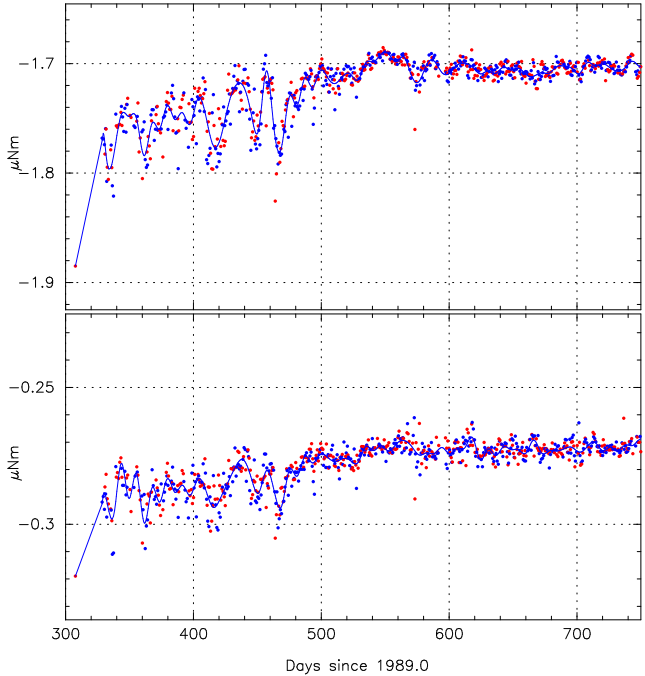}}
\caption[Z-torque disturbances]{Systematic disturbances in the two main 
components of the solar-radiation torques around the spin axis after removal
of the modulation by the Earth-Sun distance variations. The disturbances
around day 400 are also observed in other components of these torques around
the spin axis, but not around the other two axes. The disturbances re-occurred
after about 570 days, when the same alignment of the Sun, Earth and satellite 
orbit returned.}
\label{fig:sm_ztorq}
\end{figure}
The extensive testing of the FDA model in the attitude reconstruction has 
exposed details on the (non-)rigidity of the spacecraft. Non-rigid behaviour 
is most clearly shown through scan-phase discontinuities (see the accompanying 
paper), likely to be the result of discrete, small displacements of one of 
the solar panels. There are various other subtle indications of non-rigid 
behaviour. One possible example of non-rigid behaviour is observed 
from the evolution of the solar radiation torques around the spin axis over 
the mission (Fig.~\ref{fig:sm_ztorq}). At times when the satellite passed 
through perigee such that 
the normal vector to the solar panels was more or less aligned with the 
velocity vector of the satellite, and thus perpendicular to the direction of
the Earth's centre (experiencing maximum force from the outer layers of the 
Earth's atmosphere), the solar radiation torque components for the spin axis
became significantly disturbed. These torque components, which describe 
the amplitudes of typical harmonics of the rotation period of the satellite,
result from shadows of the solar panels on the spacecraft. In order
to disturb the amplitudes in a systematic and correlated manner, there are
two possibilities: variations in the solar radiation, or variations in the
surface of the satellite as seen from the Sun. Solar-flux variations are more 
than an order of magnitude smaller than the level required by the observed 
effects, and would also affect the torques on the $x$ and $y$ axes, which 
is not observed. This leaves only changes in the outer properties of the
satellite as an explanation, even though most of these would probably also be 
observed on the $x$ and $y$ axes. It is still unclear what properties of the
satellite could be changed to cause these quite significant and systematic
changes. 

\begin{figure}
\centerline{\includegraphics[width=8.5cm]{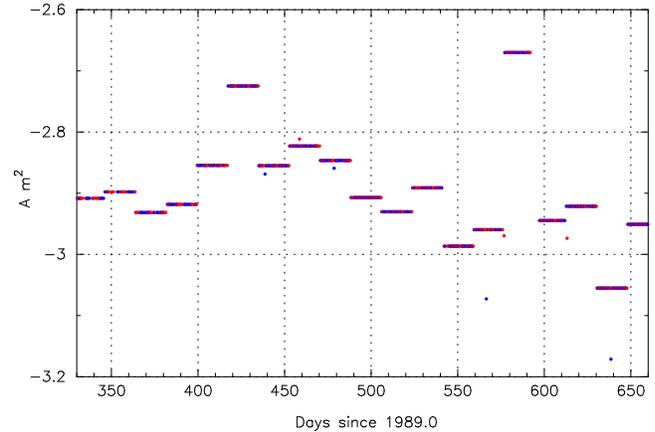}}
\caption[Magnetic moment y-axis]{The reconstructed magnetic moment for the 
$y$-axis. Significant variations took place on at least time scales of a few 
days. Determinations here cover periods of 40 orbits each (with a few individual 
determinations still left).}
\label{fig:magnmomy}
\end{figure}
The non-rigidity effects described above are probably causing significant
disturbances on the apparent rotations of the satellite, most of which cannot
easily be recognised or predicted. The result of this is that the prediction 
of the torques affecting the satellite can only provide a first approximation of 
its dynamical behaviour. The detailed behaviour has to be reconstructed for 
each orbit separately. Further 
disturbances are probably caused by unresolved small hits. Prediction of
torques is also hampered by the presence of magnetic torques, and variations
in the satellite's magnetic moment (Fig.~\ref{fig:magnmomy}) and the magnetic 
flux in the Earth's magnetic field. The magnetic moment variations may, 
occasionally, have been a source
of violation of the basic condition of the FDA by causing small discrete 
variations in the external torques, but for most of the time the underlying
torques will still have been continuous. Any magnetic torque variations will 
be more significant close to perigee (where no observations took place), 
as around apogee the magnetic field is weak and the resulting torques are 
quite small. 

\subsection{Time resolution and weights}
\begin{figure}
\centerline{\includegraphics[width=7cm]{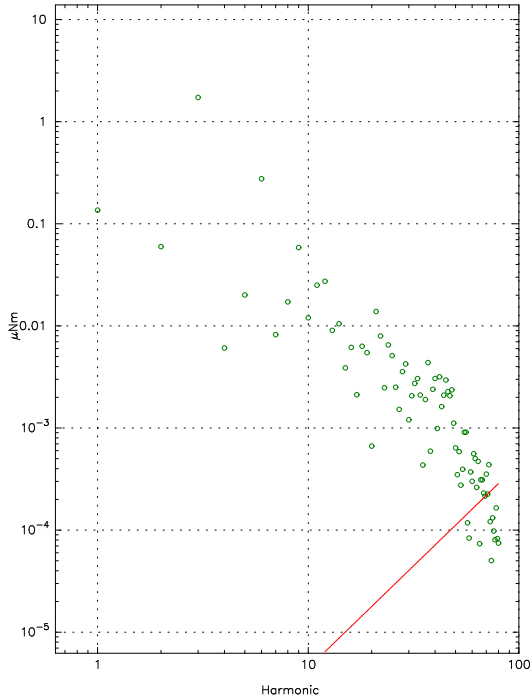}}
\caption[Power spectrum z-torques]{The power spectrum of the spin-synchronous 
harmonics in the torques acting on the spin axis. The three-fold
symmetry of the satellite reflects in the relatively higher amplitudes
for the 3$^\mathrm{rd}$, 6$^\mathrm{th}$, 9$^\mathrm{th}$ and 12$^\mathrm{th}$ 
harmonics. The line indicates values that would be equivalent to
causing a 0.3~mas amplitude in the positional variations.
 }
\label{fig:powerz}
\end{figure}
\begin{figure}
\centerline{\includegraphics[width=8.5cm]{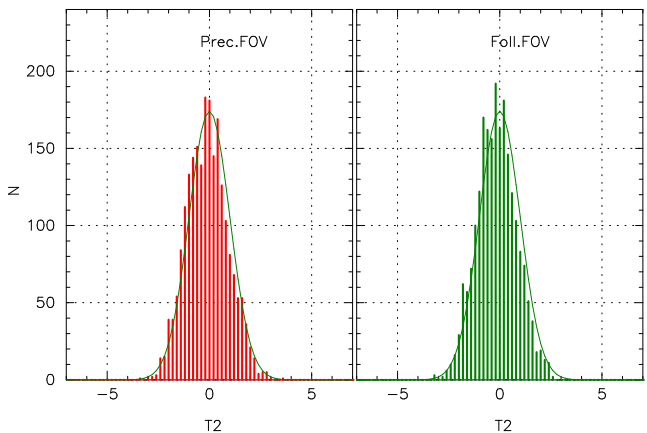}}
\caption[T2 for format abscissa]{The distribution of T2 statistics for the
construction of the mean abscissa residuals per format of 10.7~s. The curves 
show the ideal Gaussian distribution for the same number of observations.
The offsets towards lower T2 values are due to the weight limit applied to 
the brightest transits. 
 }
\label{fig:t2format}
\end{figure}

The time resolution used in the along-scan attitude reconstruction, as well
as the density of nodes in the spline fitting, has been determined in first 
instance from the power spectrum of the harmonic signals in the observed 
torques (Fig.~\ref{fig:powerz}). As a comparison, an amplitude 
$A_n=10^{-4}\mu$Nm for the $n^\mathrm{th}$ harmonic in the torques results 
in an amplitude for the positional variations of $671/n^2$~mas. 
Thus, harmonics up to about $n=70$ are still contributing significant (more 
than 0.3~mas) positional variations. A full cycle for $n=70$ is equivalent 
to about 5\degr\ on the circle, and 100~s of time, and will require two intervals
between nodes to be approximated. Nodes in the along-scan 
attitude have been placed at nominal distances of about 66~s, and the 
resolution of single observations at 10.7~s. A single observation combines,
as a weighted mean, single-star transit residuals as observed in one field of 
view. An upper limit is applied to the weight contributions of bright stars 
to restrict their influence. As a result, the T2 statistics 
\citep[see for example][]{papoul91} for these
observations are slightly offset (Fig.~\ref{fig:t2format}). Any severe 
outliers are eliminated from calculating the mean.

An important new aspect in the final along-scan attitude reconstruction is
the control over the weights of contributions from the two fields of view
as applied in the solution. As has been described in the accompanying paper,
it is essential for the production of absolute parallaxes and the rigidity
of the reconstructed sky to ensure very good connectivity between the fields 
of view. This can only be achieved if both fields of view are contributing 
significantly to the reconstruction of the along-scan attitude. In order
to obtain a good weight distribution without losing too much information, 
the weight contributions are assessed for each node interval in the 
spline function. The weight contributions from the individual observations 
are added up for each interval, and for each field of view. Whenever the 
total interval weight of one field of view exceeds that of the other field 
of view by more than a factor 2.89, the higher weight is reduced to produce 
the maximum allowed ratio. During the initial tests the weight ratio has been 
reduced from a starting value of 6 to the current value of 2.89, without 
noticeable deterioration of the abscissa residuals for the brightest stars. 
Even at a maximum ratio of 2.89 there are still many iterations needed to
obtain reliable absolute parallaxes for all stars in the catalogue. During 
these iterations minor distortions of the reconstructed sky are slowly 
smoothed out.

\section{Internal statistical tests}
\label{sec:intstat}
A very tight control on the statistical properties of the new reduction
has ensured the detection of scan-phase discontinuities, external hits and
a few other peculiarities in the data, and has prevented these from 
penetrating the final astrometric results. The control covers various 
aspects which can be split in two groups: trend analysis and analysis of 
residual distributions. In some cases combinations of the two are applied too.
 
\subsection{Trend analysis}
Trend analysis is applied to calibration parameters in order
to detect outliers that may indicate problems in the reduction. Application
to the basic-angle and instrument parameter calibrations, for example, exposes 
those data sets for which temperature control in the payload was temporarily
(partly) lost. Such data sets can either be repaired (allowing additional
modelling parameters in the reduction) or rejected.

\begin{figure}
\centerline{\includegraphics[width=8.5cm]{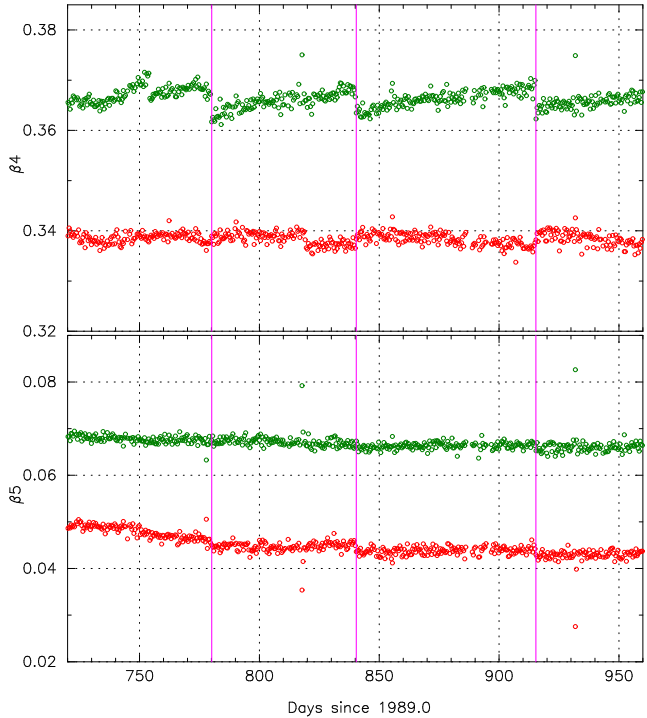}}
\caption[OTF trend analysis]{An example of trend analysis, applied here to the
calibration of the second-harmonic modulation parameters for the main 
detector, $\beta_4$ and $\beta_5$ (see also Section~\ref{sec:modul}). Next 
to the discontinuities 
caused by re-focusing (vertical lines) disturbances can be seen on day
755 (orbit 1006, change of thermal-control electronics), days 779, 856 and 932
(orbits 1060, 1235 and 1407, re-starting the on-board computer),
and day 818 (orbits 1150 \& 1151, anomalous voltage). In each graph the
upper set of data points refers to the following and the lower set 
to the preceding field of view.
 }
\label{fig:otftrend}
\end{figure}
Trend analysis has been applied to:
\begin{itemize}
\item the noise, drift and orientation properties of the gyros;
\item the geometric calibration of the star mapper grid;
\item the noise properties of the star mapper based attitude solutions;
\item the calibration of the optical transfer function for the main
detector (describing the relation between the first and second harmonics
in the modulated signal as a function of field of view, position on the grid 
and colour of the star);
\item photometric calibration parameters for the main grid;
\item instrument parameters and basic angle for the main grid;
\item the calibration of scan-rate changes as a function of thruster-firing 
lengths;
\item torque calibration.
\end{itemize}
In trend analysis the accumulated calibration parameters are displayed as a 
function of time,
with the possibility to focus in or out or pan through the data, displaying
the identification of time and orbit number for any position of the cursor 
in the graph. An example of such a graph is shown in 
Fig.~\ref{fig:otftrend} for the calibration of two modulation parameters 
of the main detector signal, $\beta_4$ and $\beta_5$ (see also 
Section~\ref{sec:modul}), which together describe
the amplitude ratio and phase difference between the first 
and second harmonics. It shows the different reaction of the two parameters
in the two fields of view to a range of events such as refocusing,
changes in thermal control, and a voltage outing. Most of these events
also show up in basic angle drifts (see accompanying paper). Thus, trend 
analysis is used to detect the occasional anomalous conditions for payload
operations that may require special attention. By applying it to a wide 
range of parameters, few, if any, anomalous conditions slip through unnoticed.

\subsection{The star mapper based attitude reconstruction}

\begin{figure}
\centerline{\includegraphics[width=8.5cm]{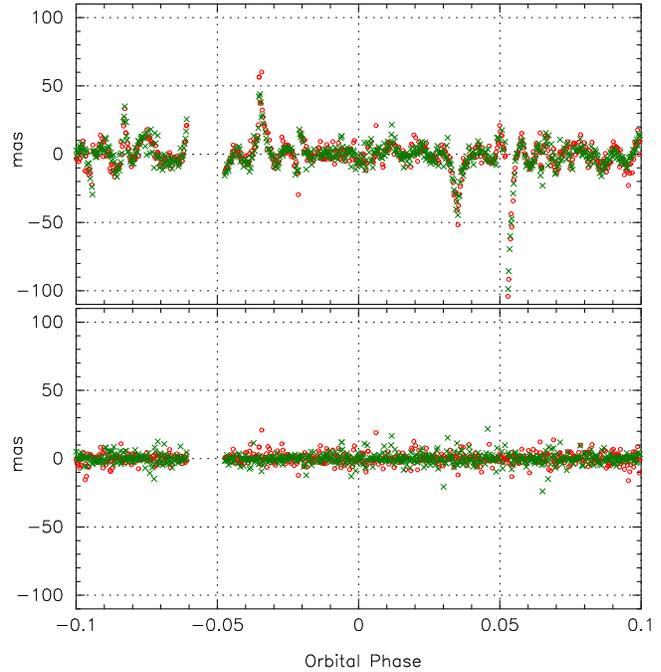}}
\caption[Abscissa residuals or0401]{Abscissa residuals for one rotation of the
satellite in orbit 401 (early May 1990). The top graph shows the residuals 
relative to the star mapper based attitude, and displays effectively the
performance of that process. The bottom graph shows the same observations 
after the final iteration in the along-scan attitude fitting. The crosses
and circles refer to observations from the preceding and following fields
of view. The discontinuities in the upper graph reflect the effect of thruster
firings. 
 }
\label{fig:abscres}
\end{figure}
The performance of the first step in building the attitude model is 
illustrated in Fig.~\ref{fig:abscres}. In the top graph is shown how the
star mapper based attitude (SMA) performs for the reconstruction of the 
scan phase. With the relatively high noise levels and low density of the 
star mapper data, the attitude detail that can be fitted using those data is
limited. The overall accuracy of the reconstruction is generally within
10~mas, but systematic excursions occur fairly frequently, and can reach 
up to 200~mas. The main reason for these excursions is generally lack of
suitable data. This applies even more so for the reconstruction of the 
spin axis position: while transits from both fields of view contribute
to the along-scan attitude, it is only one field of view that contributes to
the reconstruction of one direction of the spin axis. 

\begin{figure}
\centerline{\includegraphics[width=8.5cm]{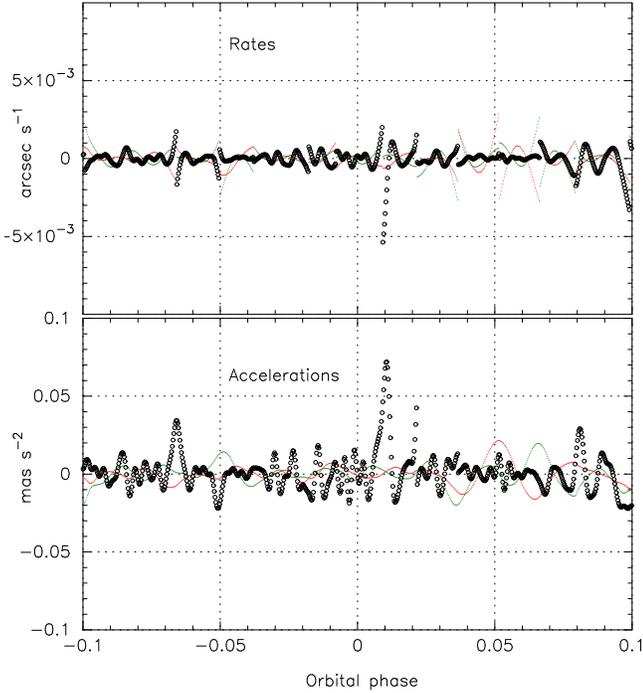}}
\caption[Rate and Acc residuals or0052]{Residuals in rates (top) and 
accelerations (bottom)
for the star mapper based attitude reconstruction. The data cover one rotation
of the satellite. The large symbols refer to the along-scan direction, the
small symbols to the spin-axis position. Data for orbit 52, 27 November 1989.
 }
\label{fig:ratacc}
\end{figure}
The SMA determines more than just the reference
pointing model. It also provides information on the inertial rates of the
satellite on all three axes. This is required for the reduction of the 
main grid transits: the rotation rates of the satellite, together with
the grid geometry, determine the relation between time and relative modulation
phase for the individual samplings of a star transit. The along-scan rate 
errors as derived from the star mapper data can be as large as 
$\pm6$~mas\persec, but are typically within $\pm1$~mas\persec\ 
(Fig.~\ref{fig:ratacc}). These errors will only cause a very small amount of 
additional noise on the amplitudes of the estimated modulation parameters. 
At 6~mas\persec\ the reconstructed amplitudes for the first harmonic is 
decreased by about 0.004~per~cent, and for the second harmonic by about 
0.016~per~cent. Acceleration errors for the SMA are generally
between $\pm0.06$~mas~s$^{-2}$ and create errors on the phase estimates 
below $10^{-3}$~mas. Errors on the SMA are thus unlikely to contribute any 
significant noise to the estimates of the modulation parameters for the 
main-grid transits.   

\subsection{The modulation parameters}
\label{sec:modul}

\begin{figure}
\centerline{\includegraphics[width=8cm]{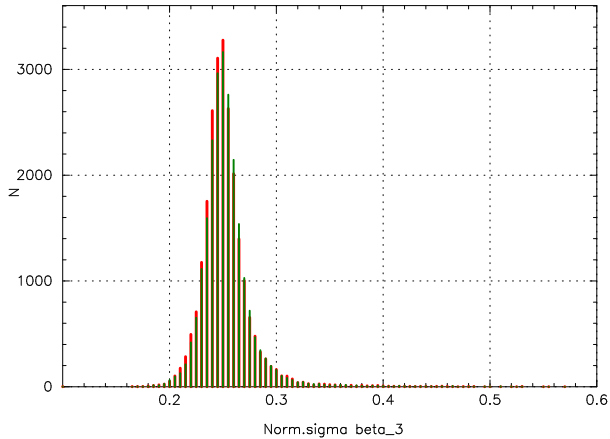}}
\caption[Modulation phase errors hist. or0075]{The distribution of formal 
errors on the modulation phase $\beta_3$ in arcsec, multiplied by the square 
root of the total photon count and the relative modulation amplitude $M1$ 
for the signal. Data for orbit 75.}
\label{fig:sb3hist}
\end{figure}
The analysis of the modulated signal obtained from transits over the main grid
is central to the Hipparcos data analysis. Here we follow the NDAC model, 
which describes the characteristics of the second harmonic relative to 
the first harmonic parameters:
\begin{eqnarray}
I_k({\bf \beta}) &=& \beta_1+\beta_2\bigl[\cos(p_k+\beta_3) \nonumber \\
&+& \beta_4\cos 
2(p_k + \beta_3) + \beta_5\sin 2(p_k + \beta_3)\bigr],
\label{eq:beta_par}
\end{eqnarray}
where $p_k$ is the assumed modulation phase of the signal $I_k$ for sample
$k$ as based on the attitude model, which can be the SMA or the result
of a previous iteration. The astrometric information is derived from 
the modulation reference phase, $\beta_3$, 
while $\beta_4$ and $\beta_5$ together describe the intensity-independent 
relative amplitude and phase of the second harmonic in the modulated signal. 
The relations between these parameters and other representations of the 
modulated signal that lend themselves better for a least-squares solution 
can be found in Volume~3 of \citet{esa97} or \citet{fvl97}.

\begin{figure}
\centerline{\includegraphics[width=8.5truecm]{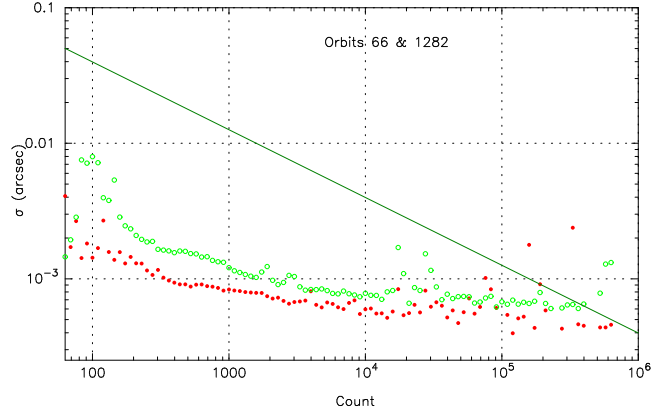}}
\caption{A log-log diagram of the mean formal errors on the predicted 
positions, as derived from the published catalogue, as a function of the 
total photon count for frame transits. The data
for orbits 66 (open circles) and 1282 (dots) are shown. The formal errors
for orbit 66, at the beginning of the mission, are clearly larger (as a result
of uncertainties in the reconstructed proper motions) than
for orbit 1282, half-way the beginning and end of the mission. The 
diagonal line shows the average
photon-noise relation over the mission for a modulation parameter $M1=0.72$.}
\label{fig:PredPosErr}
\end{figure} 
The formal error on $\beta_3$ is a function of the total photon count 
received and the relative modulation amplitude $M_1\equiv\beta_2/\beta_1$, 
which has some dependence on field of view, colour index of the star and 
position on the grid. The modulation amplitude varies as a function of time 
as a result of small changes in the optics and has a typical value of 0.72. 
Over the mission, the error on the modulation phase has been calibrated as:
\begin{equation}
\sigma\beta_3\approx \frac{255}{M1\sqrt{I_\mathrm{tot}}}~~mas,
\label{eq:sb3}
\end{equation}
where $I_\mathrm{tot}$ is the total photon count accumulated for the signal. 
The factor 255 in Eq.~\ref{eq:sb3} represents
an average over the field of view. Over the mission the average amplitude
over the field of view decreased slowly, causing this factor to evolve from
a value of 249 at the beginning of the mission to 257 for the preceding
and 253 for the following field of view. The recovery of this factor is 
illustrated for one orbit in Fig.~\ref{fig:sb3hist}.  Orbits affected by 
thermal control problems show generally lower modulation amplitudes and 
thus larger errors on the phase estimates.

\begin{figure}
\centerline{\includegraphics[width=8.5truecm]{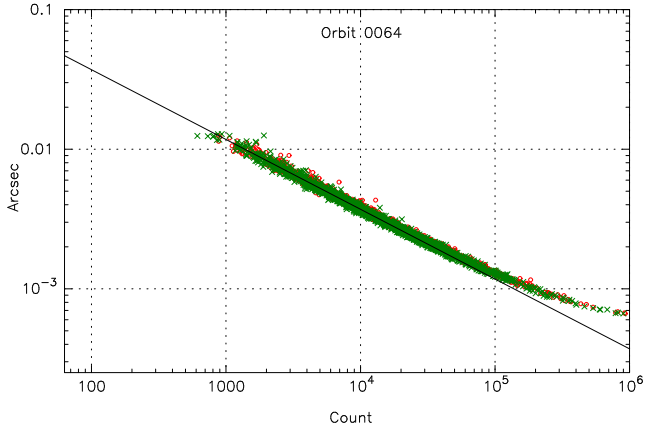}}
\caption{A log-log diagram of the formal errors on the input data for the
along-scan attitude reconstruction. Each data point represents 10.7~s of
observations in one of the fields of view. The diagonal line is the photon
noise relation, and the deviation towards the bright end reflects the formal
accuracies of the predicted positions of the stars.}
\label{fig:formatacc}
\end{figure} 
It is further interesting to note that for the transits with the highest 
photon counts the intrinsic precision obtained over a 2.13~s interval (a 
so-called frame transit) is already higher than the predicted positional 
accuracy for the observed star as derived from the published Hipparcos 
catalogue. This is shown in Fig.~\ref{fig:PredPosErr} for one orbit at the 
beginning of the mission, and one half-way. The predicted positions naturally
have higher errors for the start and end of the mission than half-way due to
uncertainties in the reconstructed proper motions.   

As was described above, the data for 10.7~s intervals (5 successive frame
transits, also referred to as one format) are combined per field of view to
provide the input data for the along-scan attitude corrections. The 
formal errors in the reference positions are fully taken into 
account when evaluating 
the formal error for each data point. Each format contains 
usually more than one observation per star, in which case the formal 
error for the predicted position is added to the mean of those observations
rather than to each observation individually. The resulting errors at the
time of the third iteration are shown for one orbit (64) in 
Fig.~\ref{fig:formatacc}. The range of formal errors on these data 
points is about a factor 30, but the error ratio allowed to ensure proper
connectivity is only 1.7. Most of the potentially high-precision
observations are therefore down weighted in their application to the 
along-scan attitude solution, depending on the data available in the 
other field of view. Still, in most cases a formal error of around 2.5~mas
is obtained per observation.

\subsection{Field transits and abscissa errors}

\begin{figure}
\centerline{\includegraphics[width=8.5truecm]{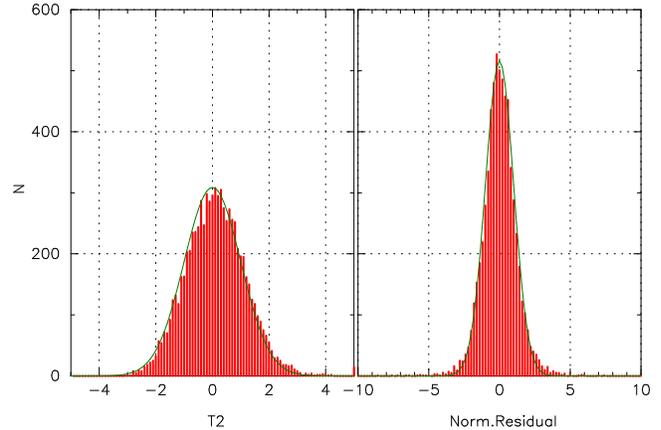}}
\caption{Statistics on field transits. Left: histogram of the T2 statistics for
the formation of field transits from frame transits. Right: normalised 
residuals between field transit abscissae and the predicted positions, for the
third iteration. The curves show the equivalent Gaussian distribution for the
same number of observations.The data are for orbit 409.}
\label{fig:fieldtrs}
\end{figure} 
An important feature of the new reduction is the time resolution of the 
final abscissa data. The Intermediate Astrometric Data in \citet{esa97} is 
in the form of one abscissa for each observed star per orbit of the satellite, 
incorporating observations from both fields of view. The great-circle 
reduction process made a higher resolution meaningless. In contrast, the 
epoch photometry for the mission is presented at field-transit level, which 
prevents a 
one-to-one relation between the photometric and astrometric information. 
In the new reduction the abscissae are preserved at field-transit level. This 
gives much improved detection possibilities for transit disturbances such as 
due to the presence of a parasitic image in the other field of view (events 
that are indicated in the epoch-photometry files). Events like these are 
characteristic for single field transits only. 

Furthermore, creating field transits provides a handle on the internal 
consistency of the data through the T2 statistics \citep{papoul91} of the 
merged frame transits, the basic 
measurement unit as described above. A typical example of the T2 statistics 
for the construction of field transits of single stars is shown in 
Fig.~\ref{fig:fieldtrs}. Also shown there is a histogram of the normalised 
residuals (weighted by the photon-noise errors only) between the field transit 
abscissae and the predicted catalogue positions (single stars) after 
the third iteration. Histograms like these are produced for each orbit 
and form part of the quality control of the data, combining checks on 
residual distributions with trend analysis.  

\section{Astrometric parameters}
\label{sec:astrompar}

The ultimate check on the internal consistency of the data comes from the 
abscissa residuals left after fitting the astrometric parameters. This
process is iteratively connected with determination and application of 
differential corrections. Some of these used to be part of the sphere 
reconstruction, but are now, due to their very small size in the new 
reductions, solved independently. These corrections are:
\begin{itemize}
\item the scan-phase zero point for each orbit,
\item small-scale distortions as a function of ordinate over the mission,
\item residual chromaticity corrections over the mission.
\end{itemize}
These corrections are described below, followed by a description of the 
statistical properties of the abscissa residuals as observed in the 
astrometric-parameter solutions.

\subsection{Final residual corrections}
\begin{figure*}
\centerline{\includegraphics[width=11truecm]{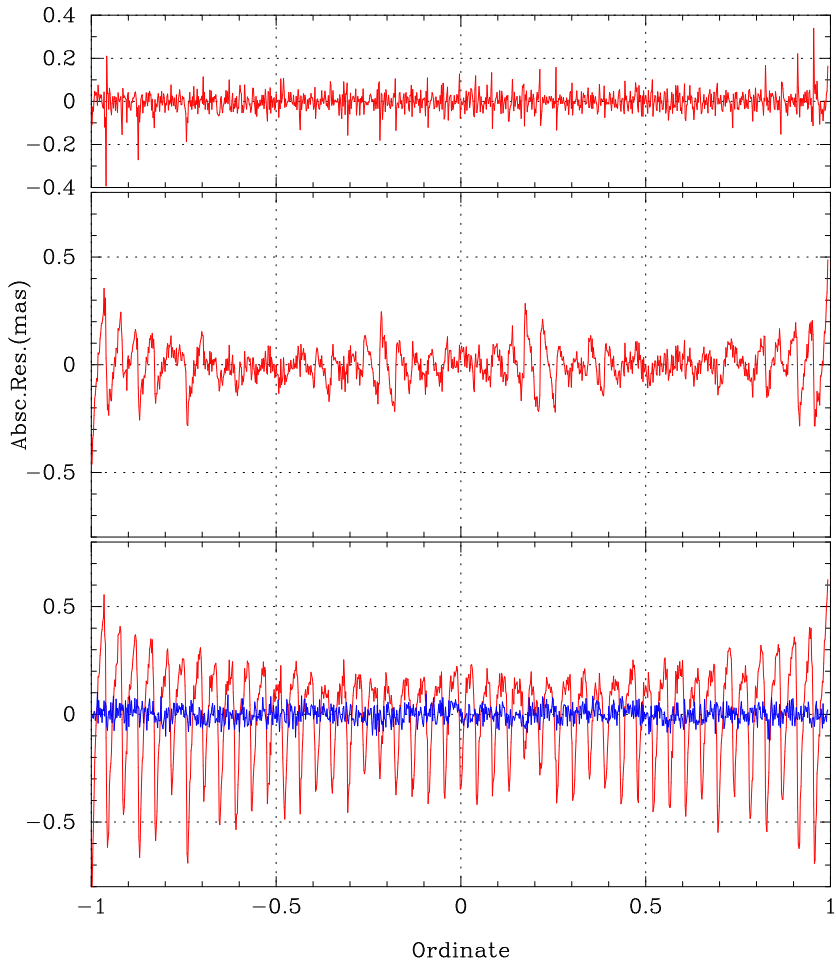}}
\caption{The small-scale distortions as 
observed across the grid for field transits. Bottom: the actual distortions, 
showing as a regular pattern the 46 individual rows of scan fields. The 
large fluctuations represent the mean over the two fields of view, the
much smaller fluctuations represent half the difference; Middle:
After correcting for a systematic non-linearity of the grid lines; Top:
After also correcting for mean scan-field tilt per row.}
\label{fig:idt_ssd}
\end{figure*} 
The scan-phase zero points, which were an important element in the original 
sphere solutions for the Hipparcos reductions, have almost lost their meaning
in the new solution. Sidestepping the great-circle solution effectively removes
the scan-phase zero point as a degree of freedom, and what is left is probably
little more than the small global distortions of the catalogue. It is therefore
no surprise that at the end of the third iteration these residuals are 
observed to be only of the order of 0.03~mas. Any orbit showing a significantly
larger value is checked for data quality. 

\begin{figure}
\centerline{\includegraphics[width=8cm]{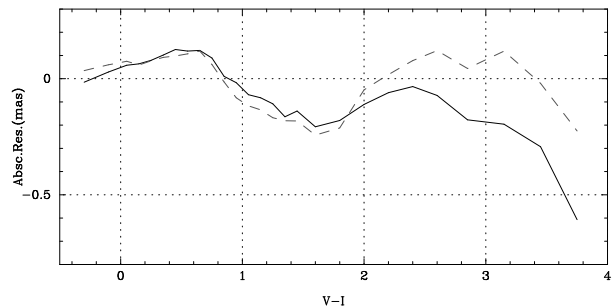}}
\caption[Chromaticity corrections]{The final chromaticity corrections
as derived from the accumulation of abscissa residuals in the astrometric
parameters solutions. The full and dashed lines refer to data from the 
preceding and following fields of view respectively. A linear term (variable 
over the mission) has already been subtracted as part of the 
instrument-parameters solution.}
\label{fig:Chrom}
\end{figure}
There are, however, small-scale distortions affecting the data at a level of
up to $\pm0.6$~mas. These originate from residual colour dependencies of the 
abscissa residuals (Fig.~\ref{fig:Chrom}) and from characteristics of the 
modulating grid, which are summarised here.

The modulating grid of 0.9 by 0.9 degrees consisted of individually 
printed scan fields, 168 along the scan-direction and 46 perpendicular to it. 
Each scan field measured about 19.28 by 70.43 arcsec  
\citep[see Fig.~5 in van Leeuwen 1997, or Fig.2.10 in Volume 2 of][]{esa97}. 
During a frame transit (at nominal scan velocity of 
168.75 arcsec\persec) a star would cross about 18 scan fields. There is, 
therefore, not much resolution in the along-scan direction. In the 
across-scan direction the resolution is determined by the tilt of the grid 
(5~arcmin) and the dispersion of the across-scan rotation rate of the 
satellite ($\approx1\arcsec$\persec). The tilt of the grid causes a 
displacement of 4.7~arcsec across-scan, while the across-scan angular-velocity 
dispersion is equivalent to about 9~arcsec in positional displacements 
in a complete field transit.

\begin{figure}
\centerline{\includegraphics[width=7cm]{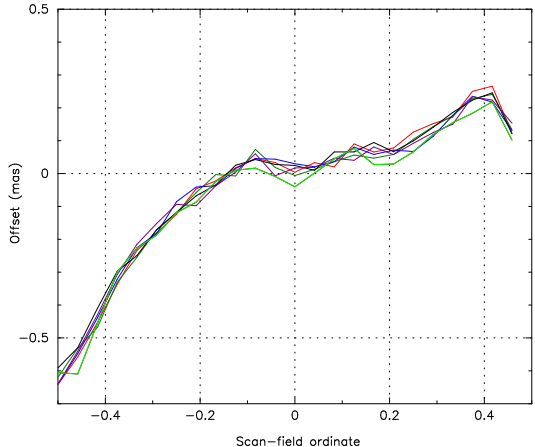}}
\caption[Grid-line curvature]{The average distortion of the grid lines
across a scan field as measured in frame transits. The seven 
curves show different intervals 
(covering each about 380 orbits) over the mission. The scan-field ordinate is
normalised to the height of a scan field, 70.43~arcsec.}
\label{fig:mssdcurv}
\end{figure}
To show the small-scale distortions as a function of transit ordinate, 
we accumulated abscissa residuals at a resolution of 24 bins per scan field, 
equivalent to a bin width of nearly 3~arcsec. The binning therefore does 
not deteriorate the actual resolution of the signal, which is set mainly by 
the across-scan angular-velocity dispersion. The bins have been carefully 
scaled and aligned with the
grid pattern, which can very easily be recognised from the accumulated 
residuals. Systematics up to 0.6~mas are exposed, revealing a general 
curvature of the grid lines with a peak-to-peak amplitude of around 
1~mas, and systematic tilts for rows of scan fields (Fig.~\ref{fig:idt_ssd}).
In a further refinement, the abscissa residuals have been accumulated at
frame-transit level, showing a systematic distortion as well as 
significant systematics in the tilting of scan fields
% as a function of the mean along-scan position of the transit 
(Fig.~\ref{fig:mssdcurv}).
In assessing these values it should be realised that on the physical grid
a scan field has a height of just under 0.5~mm, and the distortions shown
in Fig.~\ref{fig:mssdcurv} amount to displacements of up to 4~nm, compared
to an average grid period of 8.20~$\mathrm{\mu}$m.

The ordinate-dependent distortion cannot be resolved from
the published data, as the combination of field-transit data into orbit 
residuals mixes data with different ordinates. Ordinate information was 
therefore not preserved in the published data. These distortions form 
part of the unresolved modelling noise present in the FAST and NDAC 
results, but played no mayor role given the noise contribution from the
attitude modelling in those reductions. With the reduced modelling-noise 
levels of the new reduction, however, they become significant and are
fully incorporated in the instrument model. 

\subsection{Remaining abscissa dispersions}

\begin{figure}
\centerline{\includegraphics[width=7.5cm]{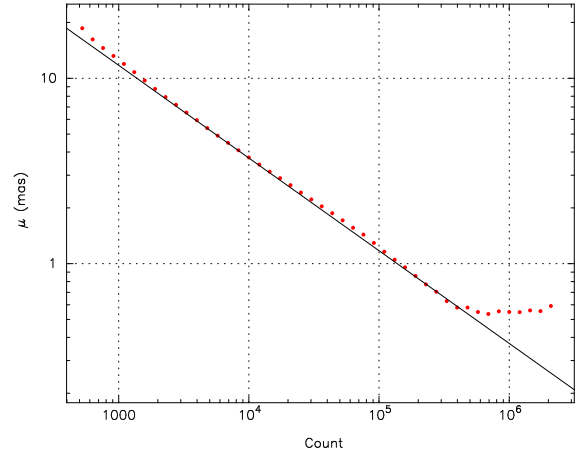}}
\caption[Abscissa dispersions]{The dispersions in field-transit abscissae 
as a function of total transit photon count. The diagonal line is the 
Poisson-noise relation for average signal modulation amplitudes.}
\label{fig:abscdisp}
\end{figure}
The dispersion of the abscissa residuals for field transits has been 
investigated as a function of the total observed photon count of the field 
transit and the modulation amplitude $M1$ (Eq.~\ref{eq:sb3}). This provides 
a direct comparison with the 
statistics observed at earlier stages of the reductions. Such a comparison 
exposed, at an earlier stage of the reductions, the presence of additional 
modelling noise, which was ultimately identified as primarily resulting from 
the scan-phase discontinuities described in the accompanying paper. After 
identifying some 1500 of these phase jumps, and incorporating that 
information in the model for the along-scan attitude reconstruction, the 
dominating noise contributions left are the Poisson and attitude noise, as 
is illustrated in Fig.~\ref{fig:abscdisp}.
The remaining attitude noise at this field-transit level is approximately
0.5~mas, about a factor 5 lower than in the original reductions, where it
is at a level of 2~mas at orbit-transit level (there are on average 3.5 to 
4 field transits for each orbit transit). The remaining Poisson noise is 
close to where one could ideally expect it to be. 

\begin{figure}
\centerline{\includegraphics[width=7.5cm]{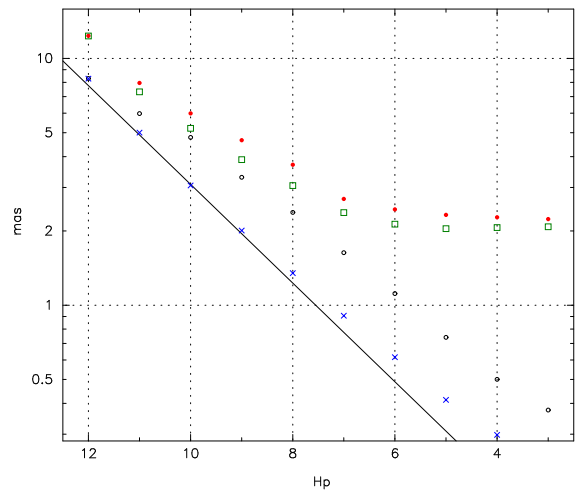}}
\caption[Fig52 new]{Dispersions in orbit-transit abscissa residuals as 
a function of magnitude. Squares: NDAC data; Filled circles: FAST data; 
Open circles: new reduction; Crosses: new reduction normalised in
observing time. The diagonal line represents the expected relation for 
photon-noise statistics.}
\label{fig:new52}
\end{figure}
In order to make a direct comparison with the published data, we combined the
field-transit abscissa residuals to create 
orbit-transit residuals. The statistics of these residuals are compared with
those of the published data as a function of magnitude in Fig.~\ref{fig:new52}.
This is the equivalent of Fig.~24 in the accompanying paper. The abscissa 
dispersions as a function of magnitude are affected by differences in 
observing time. After normalising the dispersions for average amounts of 
observing time at each magnitude, a nearly perfect Poisson relation is
recovered. The abscissa dispersions at orbit level in the new
reduction are observed to be lower than the FAST and NDAC results for all
magnitudes, but in particular for the brighter stars.

\begin{figure}
\centerline{\includegraphics[width=8.5cm]{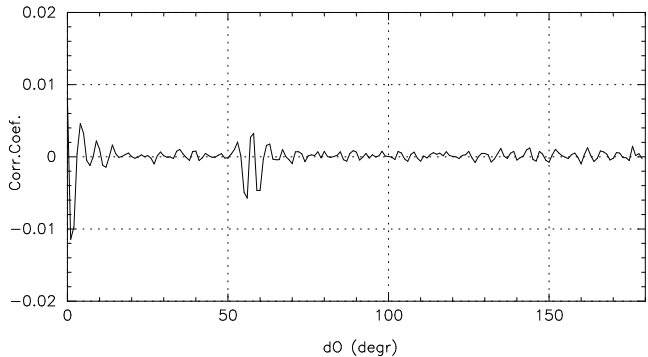}}
\caption[Abscissa error correlations]{Correlations in the abscissa errors
at orbit level for the new reduction. The various measures taken in the new 
reduction have reduced the abscissa-error correlations by a factor 30 to 40, 
to a level where they no longer play a significant role. The second set of
peaks is at the basic-angle interval of 58\degr. 
The data are for stars brighter than 7th magnitude.}
\label{fig:abscorr}
\end{figure}
A confirmation of the much lower level of attitude noise in the new reductions
is obtained from the orbit-abscissa error-correlation statistics. As is 
explained by \citet{vLDWE} and in the accompanying paper, these error 
correlations form a major obstacle in deriving reliable parallax data for 
open clusters. Figure~\ref{fig:abscorr}  shows the situation for the new 
reduction, and should be compared with a similar plot for the published 
data such as Fig.~18 in the accompanying paper. In the new reductions these 
correlations are about a factor 30 to 40 smaller, and no longer have a 
significant influence on combining data from stars in areas of a few degrees 
on the sky, thus making derivation of open-cluster parameters much simpler
and at the same time more reliable. 

\subsection{Precision of astrometric parameters}

\begin{figure}
\centerline{\includegraphics[width=8.5cm]{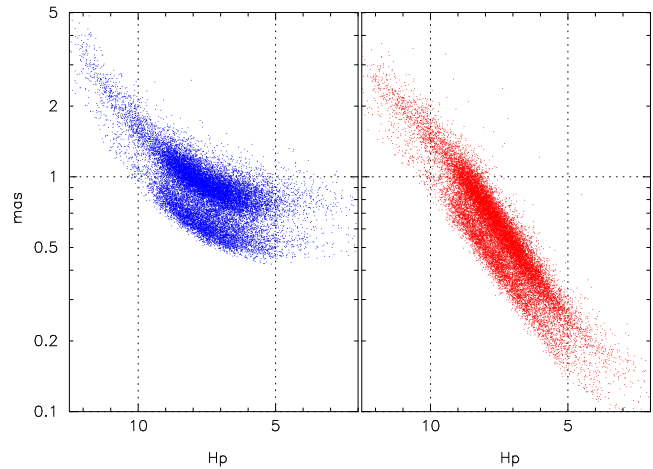}}
\caption[Precisions parallaxes]{The precisions (formal errors) of 
parallaxes in the published
data (left) and the new solution (right) as a function of magnitude. The 
bimodal structure in the plots reflects the scanning strategy: around the 
ecliptic poles the number of observations and their distribution is far more
favourable for accurate parallax measurements than around the ecliptic plane.}
\label{fig:precpar}
\end{figure}
The precision of the astrometric parameters for stars brighter than about
magnitude 7, as presented in the published data, is largely determined
by modelling noise rather than photon noise. This modelling noise has been
significantly reduced in the present study, and as expected, the same applies
to the formal errors on the astrometric parameters for the brighter stars as is
illustrated in Fig.~\ref{fig:precpar}. The main relation shown for the new 
reduction in Fig.~\ref{fig:precpar} displays the photon statistics of the 
data. The scanning strategy reflects in the two bands of formal errors:
the lower band is associated with areas within 47~degrees from the ecliptic
poles, the upper band with the area within 43 degrees from the ecliptic plane.
It is also noted that for the new reduction, unlike the published data, 
the parallax errors reflect the effects of increased observing time for 
fainter stars. This can also be observed from the relative increase
in noise for the 12$^\mathrm{th}$ magnitude stars in the published solution in
Fig.~\ref{fig:new52}. It is unclear why in the published data the errors 
for the faintest stars are larger than could be expected on the basis of 
their photon statistics.  The net result of the new reduction is an 
improvement in the formal errors over all magnitudes, and most dramatically 
so for the brighter stars. 

\section{External statistical tests}
\label{sec:extstat}
As is shown above, the new reduction has a potential parallax accuracy for 
stars brighter than Hp$=4$ of around 0.1 to 0.2~mas. At these levels of 
accuracy the tests that have been used before to verify the formal errors 
on the data have lost their effectiveness: the number of negative parallaxes 
for stars brighter than $\mathrm{Hp}=6$, for example, is significantly 
reduced (there
is only one marginal example left) and statistically irrelevant.
There are also no longer other measurements of comparable accuracy available.
However, some information can be gained from the parallax and proper motion 
data for stars in open clusters. For the Hyades the trigonometric parallaxes 
can be compared with the dynamic parallaxes. For the Pleiades, Praesepe and 
a few others the internal consistency of parallaxes and proper motions can 
be tested. These tests, which become relevant once the iterations through the
new reduction have fully converged, will be presented in one or more future
papers.

At the current stage of the reductions, the conclusion of the 7th iteration, 
the parallax adjustments can be recognised in the developments of the 
differences between the new and old solutions for stars brighter than
magnitude 3.5: these differences are still slowly increasing with every next 
iteration and are at currently nearly representative for the formal errors 
on the published data. 
This is a necessary, though not necessarily sufficient, condition for the 
formal errors on the new reduction results to be reliable. 

\section{Conclusions}
\label{sec:concl}
The new reduction of the Hipparcos data as presented here has the potential
to significantly improve overall accuracies of the astrometric data obtained
with that mission. These improvements stem from changes in the processing 
technique (the de-coupling of the great-circle reduction processes, the 
dynamical attitude modelling) and the recognition and accommodation of 
large numbers of scanning disturbances. Incorporating provisions in the
along-scan attitude reconstruction to ensure proper connectivity all over
the sky, independent of the local density and brightness of stars, ensures
a much improved reliability of the parallaxes. These provisions, however, do 
require several iterations through the reductions for convergence to be 
reached. 

\begin{acknowledgements}
It is a pleasure to thank Dafydd W.~Evans and Rudolf Le Poole for discussions 
on the subjects presented in this paper, and for reading an early version of 
the manuscript. We also thank the referee, Ulrich Bastian, for helpful 
suggestions made.  
\end{acknowledgements}

%
% BibTeX users please use
\bibliographystyle{aa}
\bibliography{3193}
% \bibliographystyle{}
% \bibliography{}

\end{document}